\documentclass[12pt,onecolumn,aps,prb,preprint]{revtex4}   

\usepackage{amsmath}    
\usepackage{amssymb}
\usepackage{graphicx}   

\begin{document}

\title{Experimental determination of heat capacities and their correlation with quantum predictions}
\author{Waqas Mahmood}

\author{Muhammad Sabieh Anwar}
\email{sabieh@lums.edu.pk}
\author{Wasif Zia}
\affiliation{School of Science and Engineering, Lahore University of Management Sciences (LUMS), Opposite Sector U, D. H. A., Lahore 54792, Pakistan.
}

\begin{abstract}
This article demonstrates an undergraduate experiment for the determination of specific heat capacities of various solids based on a calorimetric approach, where the solid vaporizes a measurable mass of liquid nitrogen. We demonstrate our technique for the metals copper and aluminum, the semi-metal graphite and also present the data in relation with Einstein's model of independent harmonic oscillators and the more accurate Debye model based on vibrational modes of a continuous crystal. Furthermore, we elucidate an interesting material property, the Verwey transition in magnetite occurring around $120$-$140~$K. We also demonstrate that the use of computer based data acquisition and subsequent statistical averaging helps reduce measurement uncertainties.
\end{abstract}

\maketitle

\section{Theoretical motivation}

The departure of the heat capacity from the classical Dulong-Petit law at low temperatures is one example of the success of quantum mechanics in describing experimental observations.\cite{kittel} The idea highlights the confluence between apparently dissimilar phenomena such as Planck's formula used in studying blackbody radiation, quantization of atomic vibrations, macroscopic heat capacities and propagation speeds of waves. An experiment that is accessible to undergraduates for measuring low temperature heat capacities and correlating results with Einstein's and Debye's quantized descriptions, is therefore, of fundamental significance. We first quickly refresh some of the underlying theory.

\subsection{Einstein's model}
The details on the specific heat capacity of solid can be looked up in any standard text on solid state physics.\cite{kittel} In summary, Dulong and Petit, using the equipartition theorem of classical thermodynamics, showed that the molar heat capacity of metals has a constant value of $3R$, $R$ being the molar gas constant. However, as we move to lower temperatures, the heat capacity shows significant variation. This deviation inspired Einstein do the first quantum mechanical calculation of the specific heat capacity.\cite{einsteinquantummechanical,simple} He assumed a solid comprising $N$ independent, three-dimensional oscillators (per unit mole), all possessing the same fundamental frequency $f$ and derived the following temperature ($T$) dependence of the molar heat capacity $C_v$,
\begin{equation}
C_v=3 N k_B \biggl(\frac{\theta_E}{T}\biggr)^2 \frac{e^{(\theta_E/T)}}{(e^{(\theta_E/T)}-1)^2},
\label{eq:einstein}
\end{equation}
where $k_B$ is Boltzmann's constant and the variable $\theta_E=hf/k_B$, with dimensions of temperature, is called the Einstein temperature. In the high temperature approximation, $T>>\theta_E$, $C_v$ approaches $3Nk_B=3R$ as predicted by the Dulong-Petit law.

\subsection{Debye's model}
Peter Debye considered~\cite{debye-original} the vibrational modes of a continuous medium, as opposed to the vibrations of independent atoms predicated in Einstein's model. The density of states $g(f)$ for a three dimensional solid is $4\pi f^2/c^3$, where $c$ is the propagation speed of the wave inside the solid. Since the speeds vary for the longitudinal ($c_l$) and the doubly-degenerate transverse ($c_t$) waves, one may also write,
\begin{equation}
g(f)\,df = 4\pi f^2 \biggl( \frac{1}{c_l^3}+\frac{2}{c_t^3}\biggr)\,df.
\end{equation}
Using the above frequency spectrum and Planck's formula, the total vibrational energy of the crystal is,
\begin{equation}\label{eq:average_energy}
E=\frac{4\pi k_B^4}{h^3} \biggl( \frac{1}{c_l^3}+\frac{2}{c_t^3}\biggr) \, T^4 \int_0^{hf_D/k_B T} \frac{x^3}{e^x-1}\, dx.
\end{equation}
Here we have made the substitution $x=hf/k_BT$. The Debye frequency $f_D$ is a cut-off value restricting the total number of modes to $3N$, similar to the atomic vibration case. This is achieved by letting,
\begin{equation}
\int_0^{f_D} g(f)\, df = 3N.
\end{equation}
The cut-off procedure yields,
\begin{equation}
f_D^3 = \frac{9N}{4\pi} \biggl( \frac{1}{c_l^3}+\frac{2}{c_t^3}\biggr)^{-1}.
\end{equation}
Often, one defines the Debye temperature $\theta_D=hf_D/k_B$, which is a useful parameter as it determines the density of oscillators $N$ (as defined earlier) and the wave propagation speeds. The former can be used to predict the mass density of the material. The upper limit in the integral, Equation~\eqref{eq:average_energy}, therefore becomes $\theta_D/T$. Differentiating with respect to temperature yields the expression,
\begin{equation}\label{eq:debye}
C_v=\frac{4\pi k_B^4}{h^3} \biggl( \frac{1}{c_l^3}+\frac{2}{c_t^3}\biggr) \, \frac{\partial}{\partial T} \biggl( T^4 \int_0^{\theta_D/T} \frac{x^3}{e^x-1}\, dx\biggr).
\end{equation}

\subsection{Numerical predictions}
Based on either Einstein's model, Equation~\eqref{eq:einstein}, or Debye's model, Equation~\eqref{eq:debye}, one can make numerical predictions of heat capacities. Einstein's model is straightforward to simulate, while a numerical integrator is required for Equation~\eqref{eq:debye}. For example, using \texttt{Mathematica}, the integral avaluates to,
\begin{equation}
\int_0^{\theta_D/T} \frac{x^3}{e^x-1}\,dx = -\frac{\pi^4}{15}-\frac{y^4}{4}+y^3\ln{(1-e^y)}+3y^2\,Li_2(e^y)-6y\,Li_3(e^y)+6\,Li_4(e^y),
\end{equation}
where $y=\theta_D/T$ and the polylogarithm function~\cite{bailey} is defined as $Li_n(z)=\Sigma_{p=1}^{\infty} z^p/ p^n$. The numerically predicted values can be graphed and compared with experimental results.

\section{The Experiment}
\subsection{An overview}
In this work, we present an experiment for the thermal physics laboratory that aims at experimentally determining low temperature heat capacities, down to $\sim 100~$K, (a) showing deviation from Dulong-Petit law, (b) a comparison with numerical predictions of Einstein's and Debye's models and (c) an estimation of $\theta_E$ and $\theta_D$. In addition, the heat capacity is directly linked to material properties and provides insight into (d) structural (re)arrangements of atomic constituents of materials. For example, the specific heat capacity changes near phase transitions,\cite{dittman} such as at the Curie point in ferromagnets.\cite{coey} It's quite an enriching experience for students to investigate these properties inside the laboratory setting.

Thompson and White have presented~\cite{latentheat} details of a beautiful experiment measuring the latent heat of vaporization of liquid nitrogen and the specific heat capacity of various metals. Their method is based on calorimetric heat exchange between the solid and liquid nitrogen. The present discussion is a straightforward extension of their work, the added feature being the determination of low temperature heat capacities  enabled by a statistical minimization of uncertainties---a lucid example of using statistics to one's advantage. We show results from experiments performed on Cu, Al, and graphite and go on to illustrate the anomaly in the heat capacity in a ferrimagnetic material at the so-called Verwey transition.

\subsection{Experimental scheme}

\begin{figure}[!h]
\begin{center}
\includegraphics[scale=0.55]{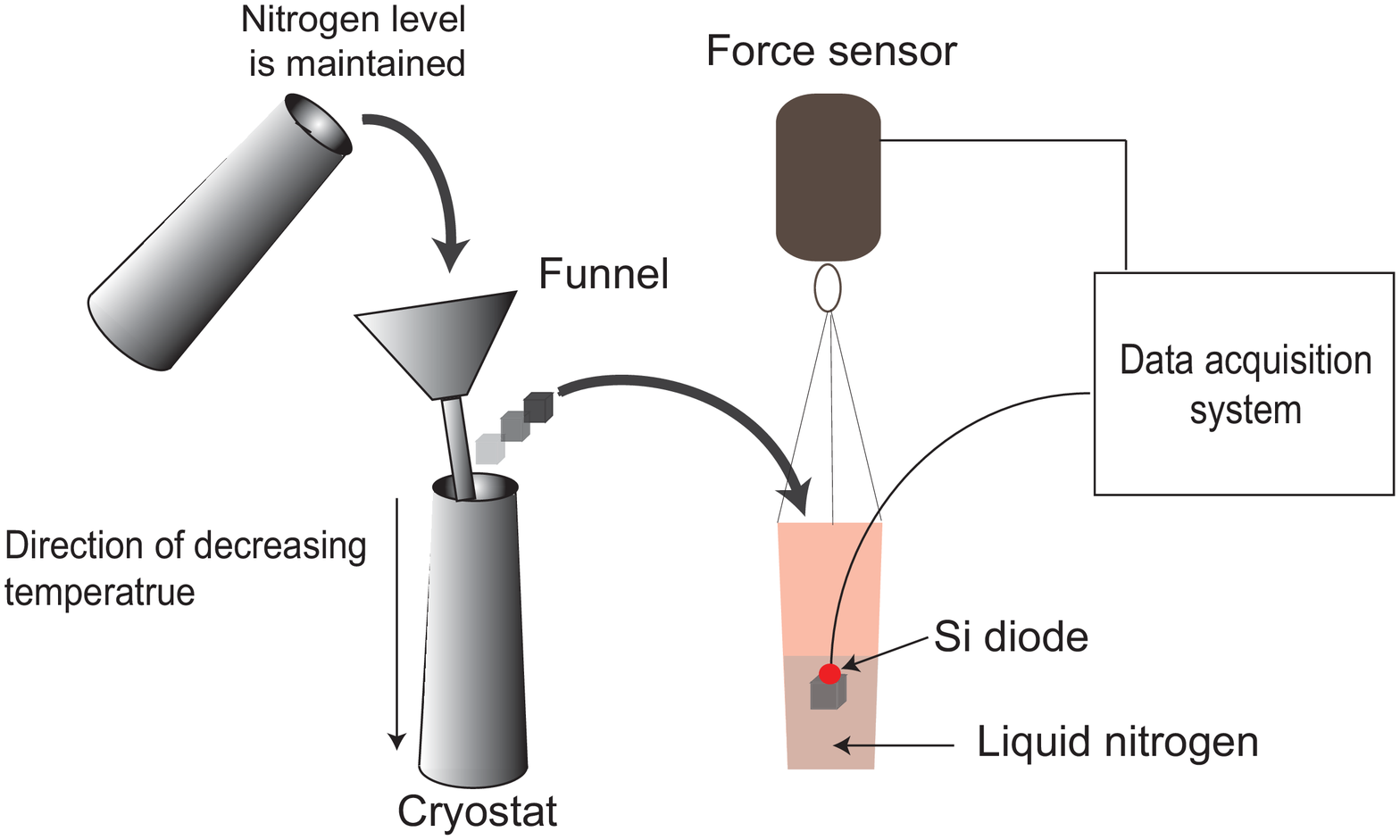}
\caption{\label{schematic} The experimental arrangement. The solid is allowed to thermally equilibrate with the cold nitrogen vapor inside a cryostat and is then swiftly migrated and dropped into a cup holding liquid nitrogen. The weight of the cup is constantly monitored while temperature is measured by passing a fixed current through and measuring the voltage drop across a Si diode.}
\end{center}
\end{figure}

\begin{figure}[h]
\begin{center}
\includegraphics[scale=0.7]{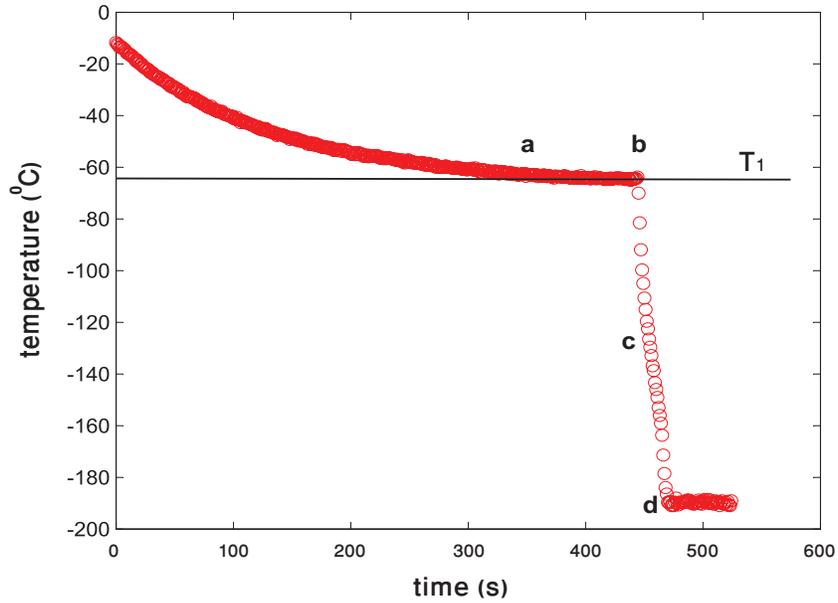}
\caption{\label{profile} Temperature profile of the solid during the course of the experiment: \textbf{a:} represents the point at which the solid achieves the desired temperature inside the vapor cryostat; \textbf{b:} shows the instant when the solid is lifted from the cryostat; \textbf{c:} is in the region of the decreasing temperature during the solid's transit from the cryostat towards and subsequent immersion into the cup, and finally \textbf{d:} is the point at which the solid eventually equilibrates at $77~$K.}
\end{center}
\end{figure}

The arrangement of the apparatus is shown in Figure~\ref{schematic}. A force sensor with an accuracy of $0.01~$N ($1~$g) (Vernier Instruments DFS-BTA) is interfaced to the computer and measures the weight (and hence mass) of a styrofoam cup containing the continuously vaporizing nitrogen. The solid, whose heat capacity is to be determined, makes contact with cold nitrogen vapor from the boiling liquid placed inside a vacuum flask which serves as our vapor cryostat. The temperature of the solid is monitored by a silicon diode that is physically secured with teflon tape while thermal grease ensures uniform thermal contact between sensor and solid. The level of the liquid nitrogen is kept constant by continuous refilling through a funnel. As the desired initial temperature of the solid, $T_1$, is achieved, it is swiftly dropped into the styrofoam cup containing liquid nitrogen. A rapid hissing sound and effervescence ensues, partly due to the Leidenfrost effect~\cite{curzon} and the cup jostles. Finally, the rapid movement settles and the background rate of vaporization is re-established. In equilibrium, the solid's temperature is $77~$K. A representative variation of the solid's temperature during the course of the experiment is illustrated in Figure~\ref{profile}.

\subsection{Measuring temperature}

We use an ordinary off-the-shelf silicon diode as our cryogenic temperature sensor.\cite{childs,precker} The underlying principle of the temperature measurement is the diode equation, $I_f=I_0 (T,E_g) \bigl(\exp{(eV_f/k_BT)}-1\bigr)$, where $I_f$ is the forward-biased current through the diode, $V_f$ is the forward-biased voltage, $k_B$ is Boltzmann's constant, $T$ is the temperature and $I_0$ is the reverse saturation current, which itself depends on temperature and the band-gap $E_g$. It can be shown~\cite{diode} that the equation relating the temperature with diode voltage is given by,
\begin{equation} \label{eq:diode}
V_f(T)= \frac{E_g}{2q}-\bigl(\log(\alpha)+\frac{3}{2}\log(T)-\log(I_f)\bigr)\frac{k_B T}{q},
\end{equation}
with,
\begin{equation}
\alpha = \frac{1}{4}\biggl(\frac{2mk_B}{\pi\hbar^2}\biggr)^\frac{3}{2}\frac{Ak_B}{\tau E_g},
\end{equation}
and $q$, $m$, $A$ and $\tau$ are, respectively, the charge and mass of the electron, cross-sectional area of the diode junction and the momentum scattering time.~\cite{kittel} A plot of $V_f$ versus $T$ is approximately linear in our temperature range of interest, \textit{i.e.}, from $\sim 100~$K to room temperature.

\begin{figure}[h]
\begin{center}
\includegraphics[scale=0.5]{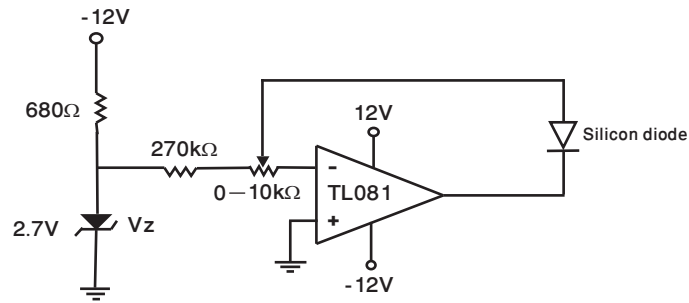}
\caption{\label{homecurrentsource} A home-built circuit for the current source to provide $10$~$\mu$A current through the silicon diode. The current is adjusted by the setting of the variable resistor.}
\end{center}
\end{figure}

To provide a constant forward biased current $I_f = 10 \ \mu$A, a current source can be easily built using an operational amplifier (TL081), Zener diode ($2.7$ V), and resistors. The circuit is shown in Figure~\ref{homecurrentsource}. The resulting voltage $V_f$ is directly read into the computer fitted with a data acquisition card (National Instruments PCI-6221) and a Labview programme automates the data acquisition.

\subsection{Why not to use a thermocouple?}

Note that unlike the wire thermocouple, the diode gives a stabler and more accurate measurement for the sample held inside the nitrogen vapor. The thermocouple operation is principled after the Seebeck effect, wherein a temperature gradient along the length of a conductor results in an emf. The thermocouple wire protruding above the liquid nitrogen surface in the cryostat is in fact placed inside a spatially extended temperature gradient. Therefore, the induced Seebeck voltage originates from the entire thermocouple wire and not only the welded tip making contact with the solid surface. Hence thermocouple measurements don't represent the temperature of the solid.

\subsection{Surface or bulk temperature?}
An additional concern that may arise is whether the surface temperature truly represents the bulk temperature. Typically our solids are cylinders of diameter $12~$mm and length $32~$mm. The Si diode indeed measures the surface temperature of the solid, but we can make an intelligent guess about the internal temperature based on the Biot number,~\cite{incropera} $Bi=HL_c/\kappa$, where $H$ is the coefficient of heat transfer in the presence of nitrogen vapor, $L_c$ is the ratio of the solid volume to its surface area and $\kappa$ is the thermal conductivity. The value of $H$ depends on the temperature of the nitrogen about its boiling point,~\cite{jin} but for a gap greater than $40~$K, it is approximately $3000~$W/(m$^2$ K). A simple calculation shows that for the Al and Cu samples, the Biot numbers are approximately $0.03$ and $0.016$, both of them are smaller than one, indicating that the temperature is uniformly distributed within the solid.

\subsection{Data analysis and statistical minimization of errors}

\begin{figure}[!h]
\begin{center}
\includegraphics[scale=0.7]{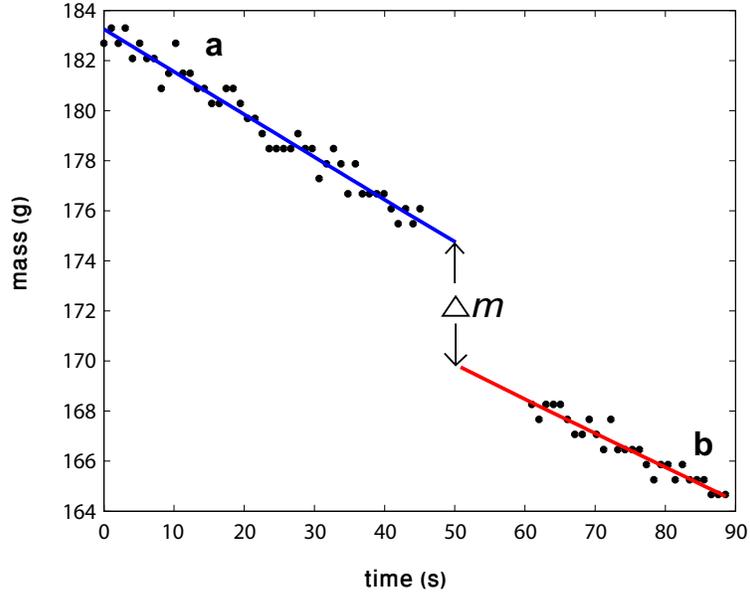}
\caption{\label{backre} Mass of the vaporizing nitrogen versus time: \textbf{a}: (blue line) background evaporation rate of boiling liquid nitrogen before immersing the solid; \textbf{b}: (red line) reestablished evaporation rate after the solid is immersed. The mass of the solid has been subtracted in plotting \textbf{b}.}
\end{center}
\end{figure}

Data from a representative experiment are shown in Figure~\ref{backre}. For the data processing part, we start off by constructing straight lines as best fits, let's call them \textbf{a} and \textbf{b} to the background evaporation rates before and after immersing the solid. The vertical displacement between these lines is a measure of the \textit{additional} nitrogen vaporized by the heat flowing from the hotter solid, $\Delta m$. The specific heat capacity is determined from the change in mass,
\begin{equation}
C_v (T_1) = \biggl(\frac{L_v \Delta m}{n_{moles} (T_1-77)}\biggr) \quad\text{J/(mol K)},
\label{eq:heatcapacity}
\end{equation}
where $n_{moles}$ is the number of moles of the solid and $L_v$ is the latent heat of vaporization of nitrogen. For the temperature dependence of $C_v$, one has to simply repeat the experiment by varying the temperature $T_1$ and calculating the corresponding decrease in mass $\Delta m$.

\begin{figure}[!h]
\begin{center}
\includegraphics[scale=0.6]{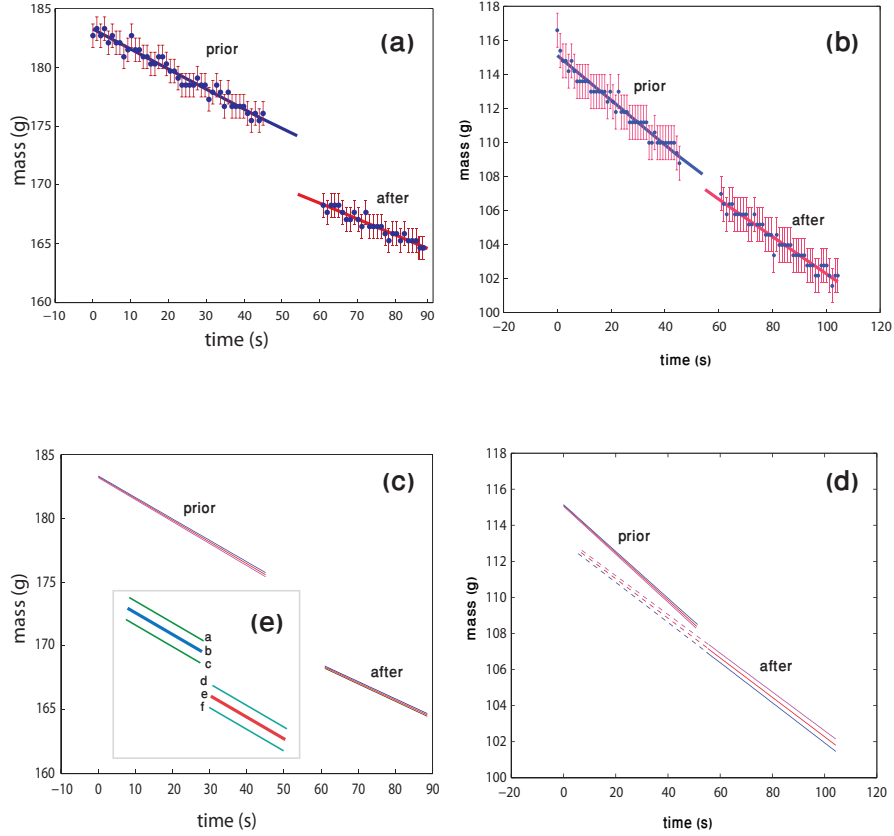}
\caption{\label{gram} (a) shows the temporal profile of the evaporating mass for $T_1=200~$K and (b) the profile for $T_1=120~$K. The time-stamped measurements of mass and the individual uncertainties in each measurement $u_m$ are presented as well as the best line fits prior and after the immersion of the solid specimen. Note that for ease of visualization, most of the sample points have been removed. (c) and (d) show the computed bands based on the uncertainties in the entire lines. The individual points and their uncertainties have been suppressed in these latter subfigures. For $T_1=120~$K, we have $\Delta m_{1} = 5.10 \ g$,  $\Delta m_{2} = 4.50 \ g$, $\Delta m = 4.80 \ g$, $u_{\Delta m} = 0.45$ \ g and $u_{C_v} = 0.44$ \ J/(mol K) and for $T_1=200~$K, we compute, $\Delta m_{1} = 1.10~$g,  $\Delta m_{2} = 0.34~$g, $\Delta m = 0.72~$g, $u_{\Delta m} = 0.43~$g and $u_{C_v} = 1.4$ \ J/(mol K). The inset (e) illustrates a close-up view of typical measurement bands in the vicinity of migrating the solid from the cryostat to the styrofoam cup. The labeling a-f is used for defining $\Delta m$, $\Delta m_{1}$ and $\Delta m_{2}$ in the text (see Equation~\ref{eq:errorcalculation}).}
\end{center}
\end{figure}

Figure~\ref{gram} shows the experimental results from solid copper, acquired at the initial temperatures $T_1$ of $200~$K (subfigures (a) and (c)) and $120~$K (subfigures (b) and (d)). The vertical lines are the error bars, let's call them $u_m$, arising from the instrumental least count, $u_m=1~$g. For a meaningful analysis, we strictly require $\Delta m > u_m$. In Figure~\ref{gram}(a), this condition is easily satisfied. The lines prior and after immersion, when extrapolated, do not statistically overlap. The situation, however, changes when the initial temperature of the solid $T_1$ decreases, resulting in smaller values of $\Delta m$. This trend is sampled in Figure~\ref{gram}(b): the instrumental uncertainties overwhelm the change in mass making it impossible to obtain a statistically reliable value of $\Delta m$.

The statistical work-around this limitation lies in appreciating that we are using the \textit{entire lines} prior and after the immersion of the solid for the determination of $\Delta m$, and not just the terminal \textit{values} at the juncture of the transferral of the solid. Therefore, we are interested in the errors in the lines, vis-a-vis, the uncertainties in the slope $u_s$ and the intercept $u_i$. These ensemble uncertainties can be computed~\cite{squires} using,
\begin{eqnarray} \label{eq:slope}
(u_s)^2 &\approx& \frac{1}{D} \frac{\sum d^2_i}{n-2},\quad\text{and}\\
\label{eq:intercept}
(u_i)^2 &\approx& \biggl(\frac{1}{n}+\frac{\bar{m}^2}{D}\biggr)\frac{\sum d^2_i}{n-2},
\end{eqnarray}
where $d_i$ is the deviation of the $i$'th experimental point from the corresponding point on the best-fit line, $\bar{m}$ is the average mass, $D$ is the sum of squares of deviations and $n$ is the total number of points in each line. Computer-based acquisition generates large amounts of data, increasing $n$ and hence reducing the uncertainties in the measurement. Based on $u_s$ and $u_i$, we can draw confidence bands, with larger $n$'s resulting in even tighter bands. These bands are illustrated in Figure~\ref{gram}(c) and (d) by a trio of lines: the middle line is the curve of best fit and the top and bottom lines represent the extrema of the band.

The best estimate for the reduction in mass is $\Delta m$ and its uncertainty $u_{\Delta m}$ is based on the maximum ($\Delta m_1$) and minimum ($\Delta m_2$) differences. These maximum and minimum are defined through the inset of Figure~\ref{gram}(e). The uncertainty is,
\begin{equation} \label{eq:errorcalculation}
(u_{\Delta m})^2= (\Delta m_1 - \Delta m)^2 + (\Delta m - \Delta m_2)^2.
\end{equation}
This work-around ensures that $u_{\Delta m} < \Delta m$ even in cases where $u_{m} > \Delta m$. A numerical example is in order here. For an initial solid temperature of $120~$K, the individual mass uncertainty $u_m=1~$g is larger than the change in mass $\Delta m=0.72~$g, but the uncertainty calculated after the statistical averaging procedure is $u_{\Delta m}=0.43~$g, which is smaller than $\Delta m$. Finally, the uncertainty in $C_v$, $u_{C_v}$ is inferred from $u_{\Delta m}$ using the well known error propagation formulas,~\cite{squires}
\begin{equation}
u_{C_v} = \frac{L_v u_{\Delta_m}}{n_{moles} ({T_1}-77)},
\label{eq:uncertainty}
\end{equation}
showing that smaller initial temperatures yield higher uncertainties $u_{C_v}$. This trend is directly observable in our results.

\section{Results and Discussion}

\subsection{Einstein fits for Cu and Al}
\begin{figure}[!h]
\begin{center}
\includegraphics[scale=0.8]{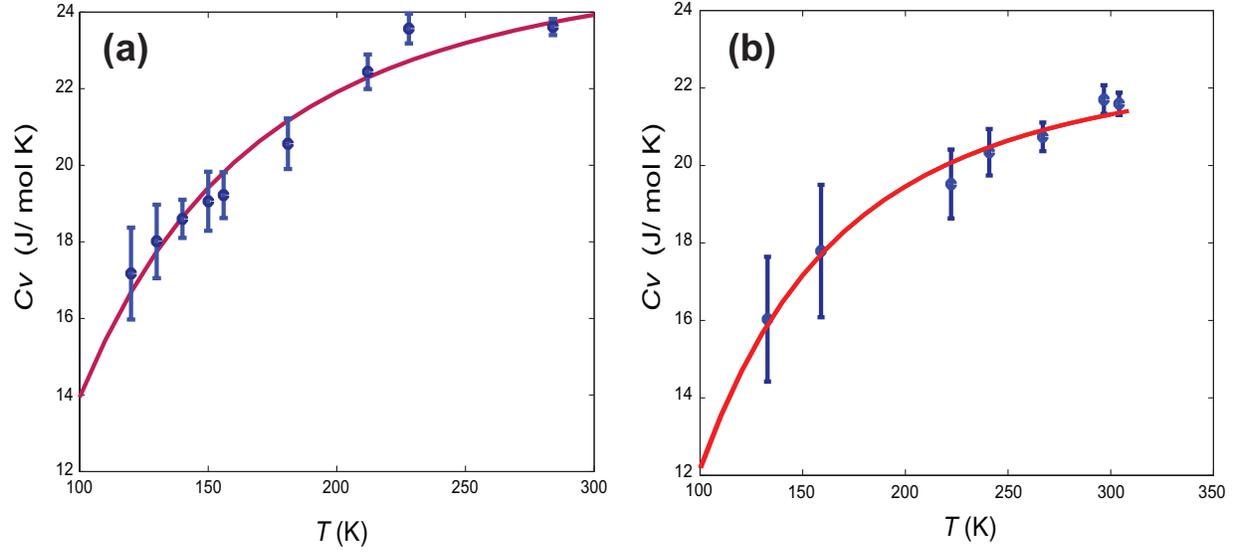}
\caption{\label{cuandal} Experimentally determined heat capacities of (a) copper and (b) aluminum along with the Einstein fits.}
\end{center}
\end{figure}

Results for the metals Cu and Al are shown in Figure~\ref{cuandal}. The data is numerically fitted to the Einstein curve, Equation~\eqref{eq:einstein}, yielding $\theta_E$'s of $278$ and $284$~K, respectively, agreeing reasonably well with the nominal values (\textit{e.g.}, $248$ and $306~$K).\cite{ledbetter} Students can also observe that the heat capacity for Al is smaller than Cu, implying higher Einstein temperatures and a higher frequency $f$. Students can get an appreciation of this fact by recognizing that the Al atom is lighter than Cu (smaller $m$) and moreover, the elastic constant is higher (higher $k$) (volume expansivity is lower). This means that for Al, the ionic frequency ($f\propto \sqrt{k/m}$) will be higher, resulting in a higher Einstein temperature $\theta_E=hf/k_B$. It is clear that the data obtained from this simple and inexpensive route show considerable agreement with qualitative predictions.

\subsection{Debye fits for Cu and Al}
\begin{figure}[!h]
\begin{center}
\includegraphics[scale=0.85]{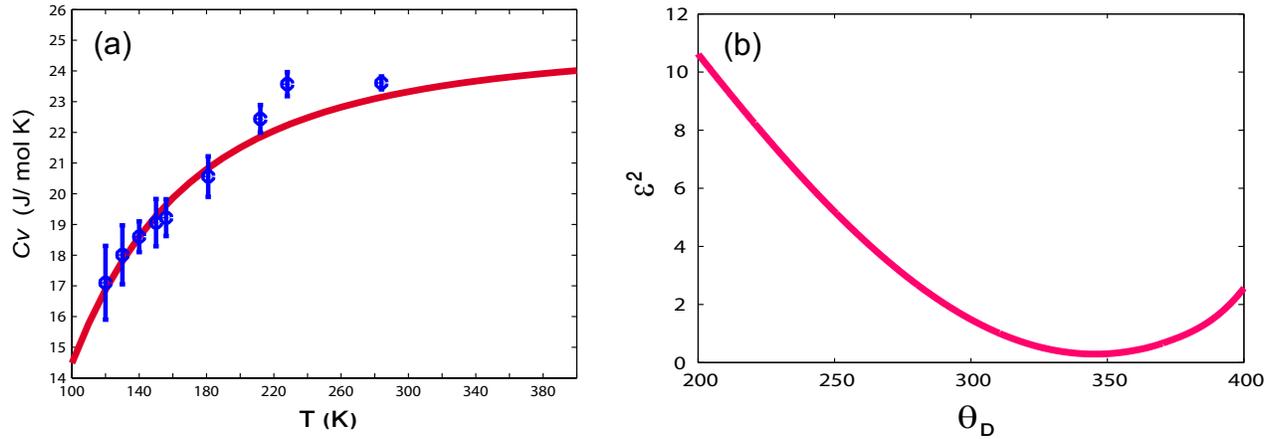}
\caption{\label{copper-debye} Experimentally determined heat capacities of (a) copper along with the best fit according to Debye's model. The fitting procedure is illustrated in (b) where the variance between the experimental points and the numerical estimates is minimized to find the best estimate of $\theta_D$.}
\end{center}
\end{figure}

The experimental data for Cu and the Debye fit with $\theta_D=350~$K is shown in Figure~\ref{copper-debye}(a), with the agreement being exceptionally good at lower temperatures. The best estimate of $\theta_D$ is found by minimizing the variance between the experimental and numerical values, $\varepsilon^2=\Sigma_p \bigl(C_v^{\textrm{(exp)}}-C_v^{\textrm{(num)}}\bigr)^2$, the process is illustrated in Figure~\ref{copper-debye}(b). Our result is in excellent agreement with the published value of $343~$K.~\cite{kittel} Analogous results from Al yield a Debye temperature of $450~$K, compared to the literature value of $428~$K.\cite{kittel}

\subsection{Heat capacity of graphite and Verwey transition in magnetite}

\begin{figure}[!h]
\begin{center}
\includegraphics[scale=0.85]{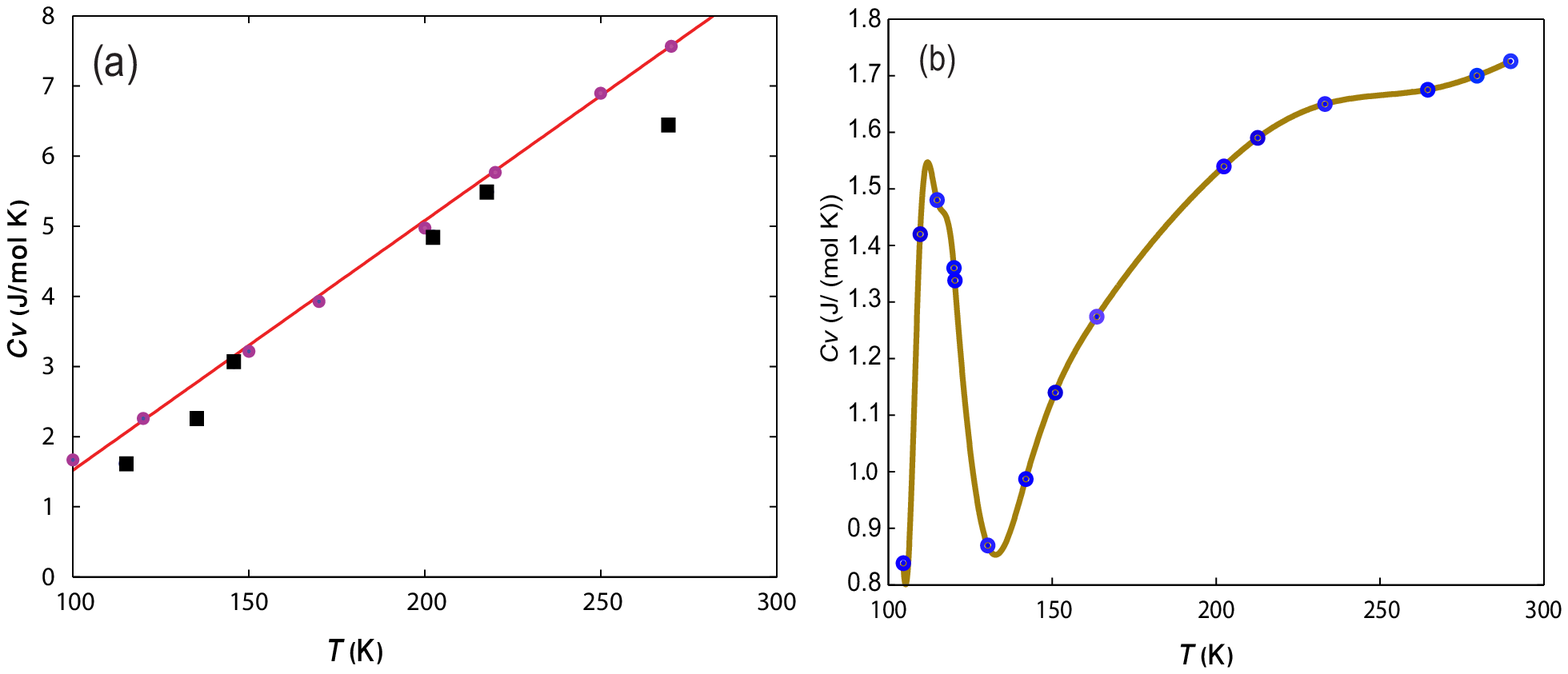}
\caption{\label{graphite-ferrite} Experimentally determined heat capacities of (a) graphite and (b) magnetite, the latter highlighting the change in specific heat capacity in the region of the Verwey transition. In (a), our experimentally determined values are shown by the squares ($\blacksquare$) and the circles ($\bullet$) are the data points from Komatsu's work.\cite{komatsu}}
\end{center}
\end{figure}

The specific heat capacity for graphite is shown in Figure~\ref{graphite-ferrite}(a) showing good agreement with published work.~\cite{komatsu} A commonly available ferrite  is magnetite, Fe$_3$O$_4$, well known for its commercial usage in ferrite cores of inductors and transformers, and its interesting magnetic, electrical and structural properties. For example, magnetite has the iconic inverse spinel structure,~\cite{cullity} important for the description of high temperature superconductivity in copper oxide based ceramics. For the present purpose, we are interested in highlighting an anomaly in magnetite's heat capacity---the Verwey transition~\cite{ferrite} which occurs around $120$-$130~$K. At temperatures above the Verwey transition, the ferrite is metallic and electrically conducting while at lower temperatures it is an electrical insulator, its conductivity decreasing by several orders of magnitude. Besides, the change in electrical conductivity, the heat capacity also shows anomalous behaviour. Our experimental results are shown in Figure~\ref{graphite-ferrite}(b) and are in agreement with Parks's original experiment~\cite{parks} in 1926.

In conclusion, the present article gives a practical illustration to determining the specific heat capacity at low temperatures using a vapor cryostat for lowering temperature, a silicon diode as a temperature probe and a gravimetric technique to track the vaporization of nitrogen with and without a solid specimen. The experimental arrangement, which is simple yet elegant, can bring home important concepts in statistical mechanics, materials physics and provide a direct demonstration of quantization of atomic or crystal vibrations. The present work builds on other foundational experiments or numerical techniques reported in the same journal.~\cite{latentheat,debye,simple} The temperature dependent measurements are easily extensible to investigating complementary material properties such as electrical conductivity and band gap,\cite{precker} as well as latent heats of phase transitions (such as the at Curie or Neel transitions and the superconducting to normal transition of high temperature cuprate superconductors). We hope this experiment will be a useful addition to the thermal physics laboratory.

\begin{thebibliography}{99}


\bibitem{kittel} C. Kittel, ``Introduction to Solid State Physics'', (John Wiley \& Sons, Delhi, 2005).

\bibitem{einsteinquantummechanical} A. Einstein, ``Planck's theory of radiation and the theory of specific heat'', Ann. Phys. 22, 180-190 (1907).

\bibitem{simple} E. Lagendijk, ``A simple determination of the Einstein temperature'', Am. J. Phys. 68 (10), 961-962, (2000).

\bibitem{debye-original} P. Debye, ``Zur theorie der specifischen w{\"{a}}rme'', Ann. Phys. 39, 789-839, (1912).

\bibitem{bailey} D. H. Bailey, P. B. Borwein and S. Plouffe, ``On the rapid computation of various polylogarithmic functions'', Math. Comp. 66, 903-913, (1997).

\bibitem{dittman} R. H. Dittman and M. W. Zemansky, ``Heat and Thermodynamics'', (Tata McGraw Hill, New Delhi, 2007).

\bibitem{coey} J. M. D. Coey, ``Magnetism and Magnetic Materials'', (Cambridge University Press, United Kingdom, 2010).

\bibitem{latentheat} C. W. Tompson and H. W. White, ``Latent heat and low-temperature heat capacity for the general physics laboratory'', Am. J. Phys. 51 (4), 362-364, (1983).

\bibitem{curzon} F. L. Curzon, ``The Leidenfrost phenomenon'', Am. J. Phys. 46, 825-828, (1978).

\bibitem{childs} P. R. N. Childs, J. R. Greenwood, and C. A. Long, ``Review of temperature measurement'', Rev. Sci. Instr. 71, 2959-2978 (2000).

\bibitem{precker} J. W. Precker and M. A. de Silva, ``Experimental estimation of the band gap in silicon and germanium
from the temperature voltage curve of diode thermometers'', Am. J. Phys. 70, 1150-1153 (2002).

\bibitem{diode} A. Khalid and M. S. Anwar, ``Superconducting Quantum Interference Devices''. This technical monograph can be downaloaded from http://physlab.lums.eud.pk//index.php/Experiments\_in\_Lab-II.

\bibitem{incropera} F.P. Incropera and D.P. Dewitt, ``Fundamentals of Heat and Mass Transfer'', (Wiley, New Jersey, 2001).

\bibitem{jin} T. Jin, J. Hong, H. Zheng, K. Tang, and Z. Gan, ``Measurement of boiling heat transfer coefficient in liquid nitrogen
bath by inverse heat conduction method'', J. Zhejiang Univ. Science A 10, 691-696 (2009).

\bibitem{squires} G. L. Squires, ``Practical Physics'', (Cambridge University Press: New York).

\bibitem{ledbetter} H. Ledbetter, ``Thermal expansion and elastic constants'', Int. J. Thermophys. 12, 637 (1991).

\bibitem{komatsu} K. Komatsu and T. Nagamiya, ``Theory of the specific heat of graphite'', J. Phys. Soc. Jpn. 6, 438-444 (1951).

\bibitem{cullity} B.D. Cullity and C.D. Graham, ``Introduction to Magnetic Materials'', (IEEE Press and Wiley, New Jersey, 2009).

\bibitem{ferrite} J. Garc{\'i}ýa and G. Sub{\'i}as, ``The Verwey transition---a new perspective'', Condens. Matter J. Phys. 16, R145--R178 (2004).

\bibitem{parks} G. S. Parks and K. K. Kelley, ``The heat capacities of some metallic oxides'', J. Phys. Chem. 30, 47--55 (1926).

\bibitem{debye} Christropher G. Deacon, John R. de Bruyn, and J. P. Whitehead, ``A simple method of determining Debye temperatures'', Am. J. Phys. 60 (5), 422-425, (1992).

\end{thebibliography}
\end{document}